\documentclass{ws-rv961x669}
\usepackage{ws-rv-van}     
\usepackage{ws-rv-thm}     
\usepackage{graphicx}
\usepackage{dcolumn}
\usepackage{bm}
\usepackage{ulem,color}

\makeindex

\begin{document}

\chapter[Jet Gravitational Waves]{Jet Gravitational Waves\label{ra_ch1}}

\author[T. Piran]{Tsvi Piran}

\address{Racah Institute for Physics,\\
The hebrew university, Jerusalem, 91904, Israel, \\
tsvi.piran@mail.huji.ac.il}

\newcommand{\tp}[1]{\textcolor{blue}{\bf #1}}
\newcommand{\rtp}[2] {{\textcolor{red}{\sout{#1}}}{\textcolor{blue}{#2}}}
\newcommand {\dtp}[1] {\textcolor{red}{\sout{#1}}}
\newcommand{\ntp}[1]{\textcolor{cyan}{[TP: #1]}}
\newcommand{\cE}{{\cal E}}
\newcommand{\apjl}{ApJL}
\newcommand{\apj}{ApJ}
\newcommand{\nat}{Nature}
\newcommand{\Eprint}{Eprint}
\newcommand{\mnras}{MNRAS}
\newcommand{\physrep}{Phys. Reports}
\newcommand{\prl}{Phys. Rev. Lett. }
\newcommand{\prd}{Phys. Rev. D. }

\begin{abstract}
The acceleration of a jet to relativistic velocities produces a unique memory type gravitational waves (GW) signal: {\it Jet-GW}. I discuss here resent result concerning properties of these GWs and consider their detectability in current and proposed detectors. Classical sources are long and short Gamma-ray bursts as well as hidden jets in core-collapse supernovae. Detection of jet-GWs from these sources will require detectors, such as the proposed BBO, DECIGO and lunar based detectors, that will operate in the deciHz band. The current LVK detectors could detect jet-GWs from a Galactic SGR flare if it is  sufficiently asymmetric. Once detected these signals could reveal information concerning jet acceleration and collimation that cannot be explored otherwise. 
\end{abstract}


\section{Introduction}
\label{sec:introduction}
The detection of gravitational waves from merging compact objects (black holes and  neutron stars) binaries opened a new era in Physics and Astronomy \cite{Abbott2015,Abbott2017GW}.  The current detectors have been upgraded and new detectors at other wavelengths are being planned. On the theory side new prospects for detection and utilization of gravitational waves from different sources have been proposed. One such less explored source - {\it gravitational waves from relativistic jets: Jet-GWs}.  

\begin{figure}
\centering
\centerline{\includegraphics[width=0.7\textwidth]{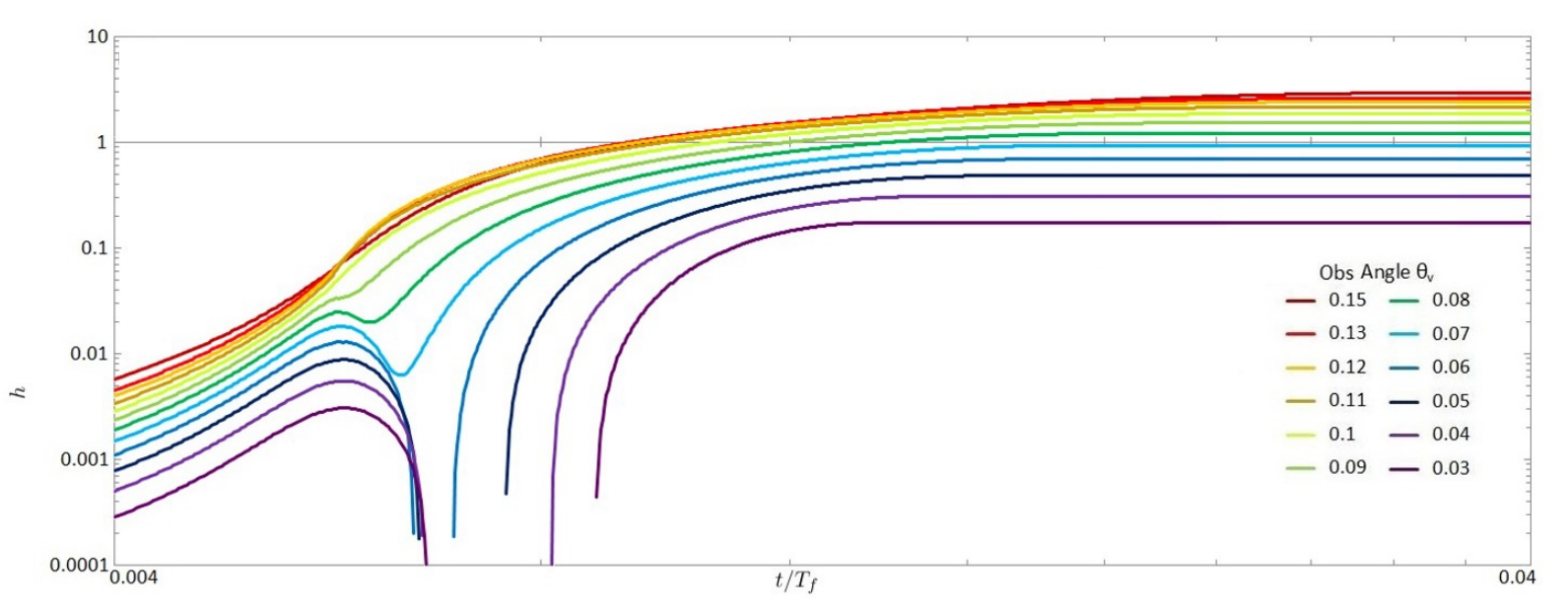}}
\caption{The normalized GW ampliture, $h(t)$, for a jet with an opening angle $\theta_{\rm j} =0.1$ and a final Lorentz factor, $\Gamma=100$ for different observer angles. From ref. [6].}
\label{fig:h1}
\end{figure}

Any relativistic jet involve acceleration of mass (or energy density - if the jet is Poynting flux dominated) from rest to a speed approaching the speed of light. This acceleration process 
produces a memory type gravitational waves signal  \cite{segalis,piran2002,Sago,ofek,Leiderschneider} (see Fig. \ref{fig:h1}). A low frequency detector, will ``see" this signal  as a sudden jump. A higher frequency  whose frequency range is comparable to the rise time of this signal, will resolve a temporal structure that    
can  reveal  the nature of the jets and  the acceleration process. In this article I review the basic features of {\it Jet-GWs} and discuss potential sources and detection prospects. 

\section{Instantaneously accelerate point particle}
\begin{figure}
\centerline{\centerline{\includegraphics[width=0.25\textwidth]{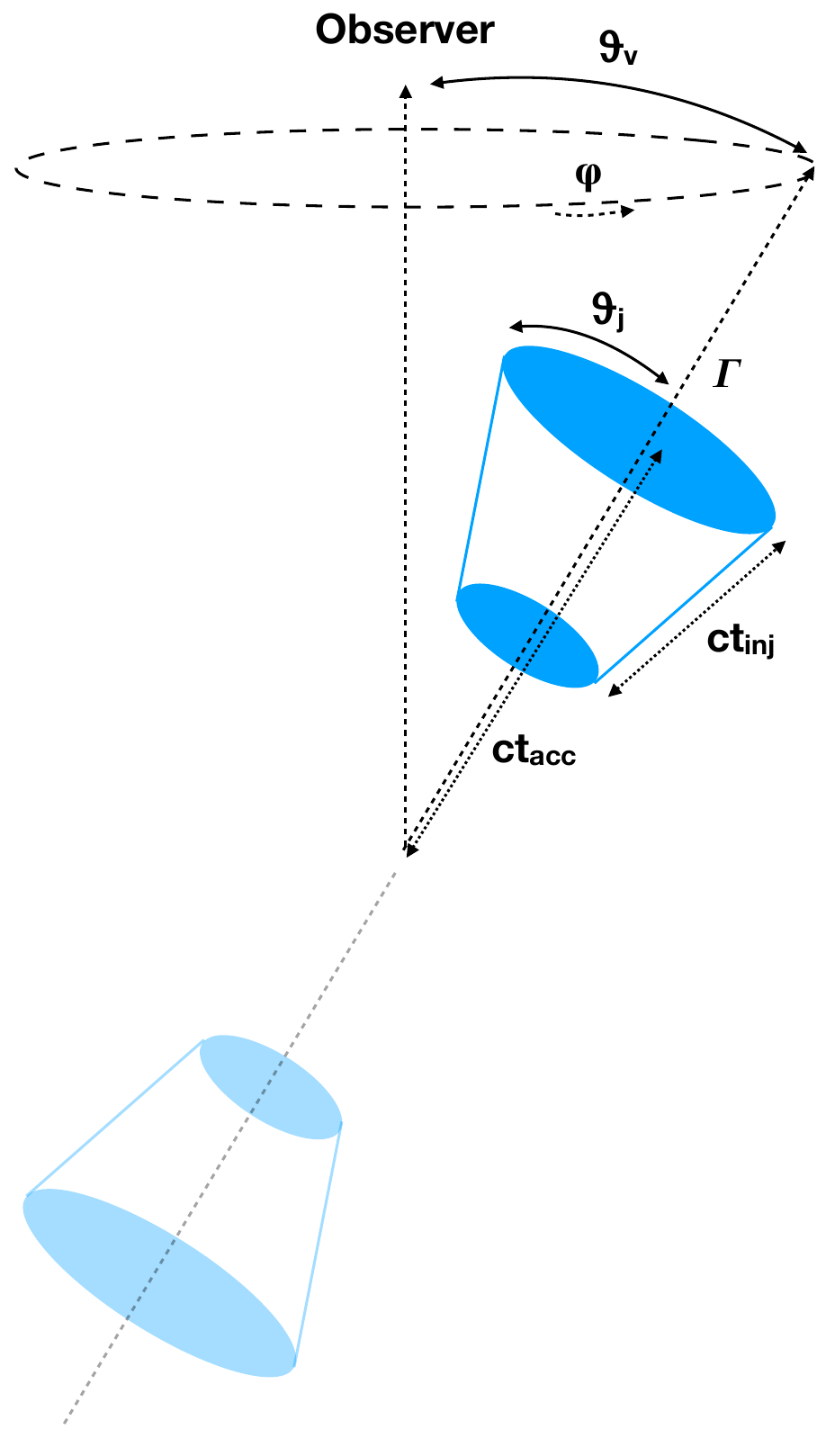}}}
\caption{A schematic description of the jet. The top shell has reached the final Lorentz factor at a distance $c t_{\rm acc}$ from the origin. The duration of mass injection is $t_{\rm inj}$. A counter-jet is shown in light colors. From ref.[7].}
\label{fig:schematic}
\end{figure}

Consider, first,    a point mass  $m$ that is instantaneously accelerated to a Lorentz factor $\Gamma$. The total energy is ${\cal E} = m \Gamma$ (unless specified otherwise I use units in which c=G=1). 
While non-physical, this limit gives an excellent idea of the emerging patterns. 
The waveform is a Heaviside step function:
\begin{equation}
\label{eq:theta}
    h(t, \theta_{\rm v}) = h(\theta_{\rm v}) \mathcal H (t) \ ,
\end{equation}  
where  $\theta_{\rm v}$ is the angle between the direction of the 
motion of the particle and the line of sight to the distant observer in the observer's frame of reference (see Fig. \ref{fig:schematic}).

The Fourier transform is given by 
\begin{equation}
\tilde h(f , \theta_{\rm v}) ={h(\theta_{\rm v})}/{f} \ .
\label{eq.FT}
\end{equation} 
The gravitational wave amplitudes $h_{\rm +}$ and  $h_{\rm x}$ of  the two polarization modes are given by \cite{segalis}:
\begin{equation}
\label{eq:phase}
    h^{TT} (\theta_{\rm v}) = h_{\rm +} + i h_{\rm x} = \frac{2 \cE \beta^2}{r} \frac{\sin^2 \theta_{\rm v}}{1-\beta \cos\theta_{\rm v}} e^{2 i \phi} \ , 
\end{equation}
where $\phi$ is an azimuthal angle. 
For a single point-particle, the phase, $2 i\phi$, can be ignored. When discussing  the metric perturbation of an ensemble of particles the complex phase  leads to a interference and must be taken into account.

\begin{figure}
\centering
\centerline{\includegraphics[width=0.5\textwidth]{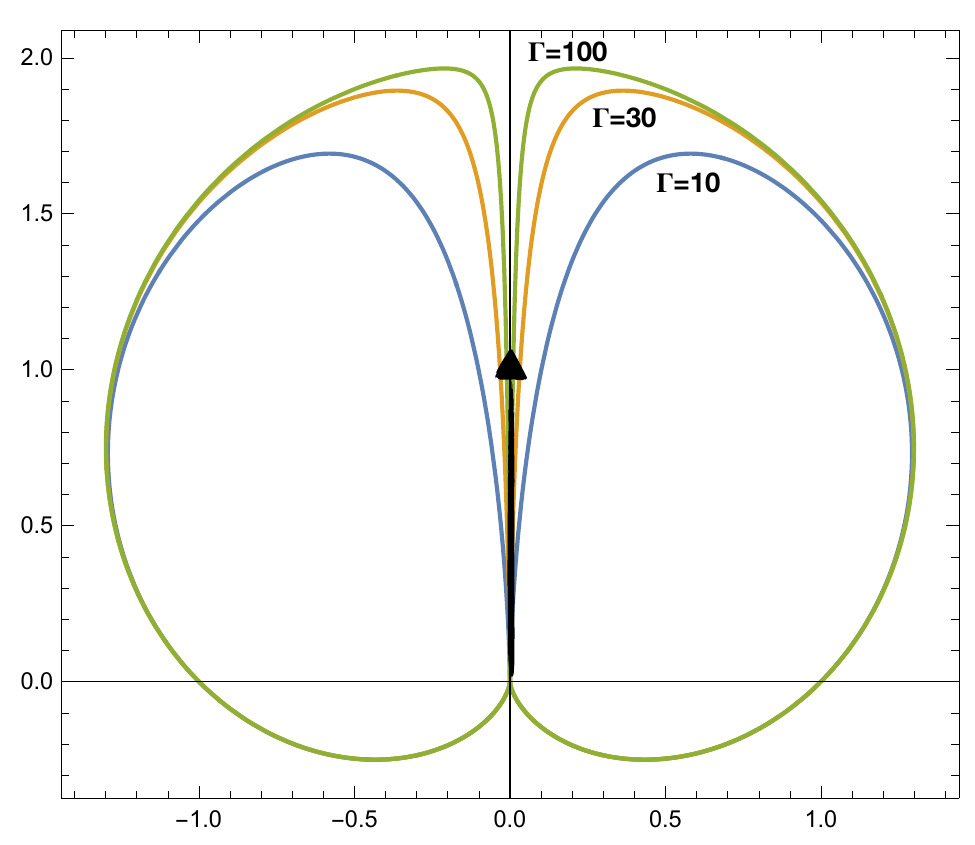}}
\caption{An antenna pattern for the GW amplitude, h, from an instantenously accelerated point-like jet for different Lorentz factors.  $h$ is anti-beamed with an opening angle $\Gamma^{-1}$ around the jet direction. }
\label{fig:de_dtheta}
\end{figure}

The angular dependence  of  $h(\theta_{\rm v})$ is shown in Fig. \ref{fig:de_dtheta}. It exhibits 
anti-beaming: the GW amplitude vanishes along its direction of motion. It remains small at a cone around it, reaching  50\% of the maximal values at an opening angle 
$\Gamma^{-1}$. The function $h(\theta_{\rm v})$ attains a maximum of
\begin{equation}
    h_{\rm max} = \frac{4{G \cE} }{c^4 r} \frac{\Gamma}{\Gamma+1},   \ \  {\rm at } \ \ \theta_{\rm max}=
    \cos^{-1}[\frac{1+\Gamma}{\beta \Gamma}] \approx 
   \sqrt{\frac{2}{\Gamma}}  \ , 
   \label{eq:hmax}
\end{equation}
where $r$ is the distance to the source and for clarify Newton's constant, $G$, and the speed of light, $c$, have been added to this equation. Interestingly,   this relation resembles the classical quadrupole relation for the GW amplitude $h_{\rm max} \propto (G M /c^2 r) (v/c)^2 $ (with $v =  c$), even though  the quadrupole approximation is not valid in this case. 

The GW energy  is  beamed in the forward direction (see Fig. \ref{fig:de_dtheta}).
50\% of the GW energy is deposited in a cone with an opening angle $\theta_{50\%} = \sqrt{{2}/{\Gamma}}$. 

\begin{figure}[h!]
\centerline{\includegraphics[width=0.5\textwidth]{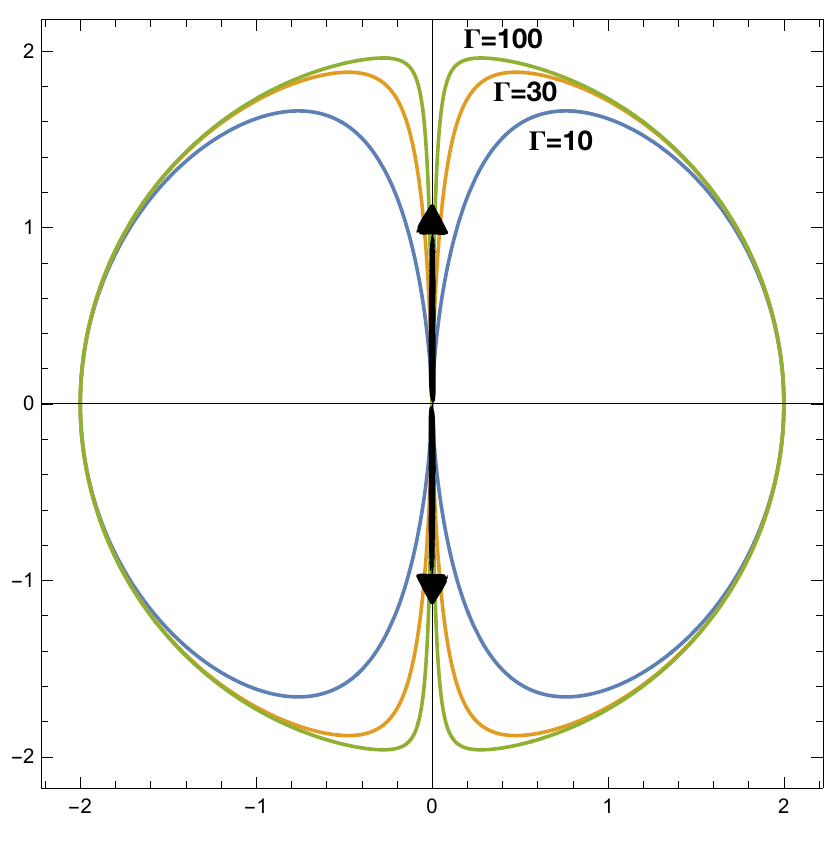}}
\caption{An antenna pattern for the GW amplitude, h, from two opposite jets for different Lorentz factors.  h is almost uniform, apart from being  anti-beamed with an opening angle $\Gamma^{-1}$ around the jets directions. }
\label{fig:twoJets}
\end{figure}

In many cases there are two jets moving in opposite directions. The GW amplitude in this case is the sum of the two signals:
\begin{equation}
\label{eq:pha}
    h (\theta_{\rm v}) =  \frac{4 \cE \beta^2}{r} \frac{ 1-\cos^2 \theta_{\rm v}}{1-\beta^2 \cos\theta_{\rm v}^2} \ ,
\end{equation}
(see Fig. \ref{fig:twoJets}). This signal is almost flat apart from two narrow {\it anti-beams} of width $\Gamma^{-1}$ along the directions of the jets. The energy is, however, still beamed into two regions of width $\sqrt{2/\Gamma} $ along the jets. Note, that these figures, that the depicts $h(\theta_v)$  are somewhat misleading. While they depict the correct value of $h$ they ignore a critical factor -   the characteristic time scale. This time scale that strongly influences the detectability (as well as the energy flux) may varies  with $\theta_v$.   

Recalling Eq. \ref{eq.FT} we note that formally the total  GW energy emitted,  
\begin{equation} 
\label{eq:egw}
E_{\rm GW} = \frac{1}{32 \pi} \iint (\frac{\partial  h}{\partial t})^2 dt d\Omega =
    \frac{1}{32 \pi} \iint   \tilde h(f)^2 f^2 d f d\Omega \ ,
\end{equation}
diverges for an instantaneously accelerating particle. This divergence is not physical as it arises from the  instantaneous  approximation. If the system has a typical timescale $t_{\rm c}$ than there should be an upper cutoff to the frequency above which $h(f)$ decreases with $f$ much faster than $f^{-1}$ and the integral is finite. 

The overall energy emitted is of order \cite{Leiderschneider}: 
\begin{equation} 
E_{\rm GW} = F(\Gamma)  [G \cE /c^5 t_{\rm c}] \cE  \ ,
\end{equation}
where $F(\Gamma)$ is a function of the jet's final Lorentz boost.
$F(\Gamma)$  vanishes when $\beta \rightarrow 0$. When $\Gamma \rightarrow \infty$ 
the situation is more complicated and in extreme cases we have to take into account the details of the acceleration process to properly estimate the energy emitted. Note that the factor in square brackets is always smaller than unity.

\section{Time Scales and the Temporal Structure}

The above considerations stress the importance of the  characteristic time scale. There are two  timescales in the system and the relevant one is of course the longest.  The two intrinsic time scales are the   injection timescale,$t_{\rm inj} $ that is measured in the observer's rest frame (which is the same as the source rest frame).  It characterizes the  overall duration of the jet. 
A second intrinsic time scale, $t_{\rm acc} $, characterizes the acceleration (it is also measured in the source rest frame).

The  signal is determined  by $t_{\rm acc}$ via  the arrival time from different angular regions of the jet. The arrival time  is related to $t_{\rm acc}$ as:
\begin{equation}
     t_{\rm o}(\theta_{\rm v}) = (1-\beta \cos\Delta \theta_{\rm v}) t_{\rm acc} ,
    \label{eq:t_obs}
\end{equation}
where $\beta$ is the jet's velocity and $\Delta \theta_{\rm v} = \theta_{\rm v}+\theta_{\rm j} $ is the sum of   $\theta_{\rm v}$ and the angular width of the jet, $\theta_{\rm j}$ (see fig:geometry). $ t_{\rm o}$ is the time difference from the observation of the beginning of the acceleration and  the the arrival time of the signal from the furthest point of the jet from observer (see Fig. \ref{fig:schematic}).

The observed time scale of the GW signal, $t_{\rm c}$,  is the longest between $t_{\rm inj} $ and $t_{\rm o}$:
\begin{equation}
t_{\rm c}(\theta_{\rm v}) = \max(t_{\rm inj},t_{\rm o}) = \max [t_{\rm inj},(1-\beta \cos\Delta \theta_{\rm v}) t_{\rm acc})] \ .
\label{eq.to}
\end{equation}
As  $ t_{\rm o}$ depends on the viewing angle, the dominant time scale may be $t_{\rm inj} $ for some observers and $t_{\rm o}$ for others. More specifically, $t_{\rm inj}$ is independent of the viewing angle. On the other hand 
$t_{\rm o} \ll t_{\rm acc}$  for small viewing angles. Thus unless $t_{\rm acc} $ is very large, $ t_{\rm c} = t_{\rm inj}$ for these angles. However $t_{\rm o}$ becomes significantly larger when $\Delta \theta$ increases and for those angles we may have $ t_{\rm c} = t_{\rm o}$

The   Fourier transform of a memory-type signal that is rising to an asymptotic value $h_0 (\theta_{\rm v})$ over a  timescale $t_{\rm c} $, has a general shape:
\begin{equation}
    \label{eq:S}
    \tilde h(f,\theta_{\rm v}) =  \begin{cases}  {h_0(\theta_{\rm v})}/{f} , & f \leq f_{\rm c}  
    \\ {h_0(\theta_{\rm v}) {g(f/f_{\rm c})} / f_{\rm c} } ,\ \ & f \geq f_{\rm c}  \ . \end{cases}
\end{equation}
 The crossover frequency, $f_{\rm c}$, in which the Fourier transform of the GW signal changes its behavior is the reciprocal of the time scale, $t_{\rm c}$:  $f_{\rm c} \equiv 1/t_{\rm c}$. The function $g(f/f_{\rm c})$ depends on the nature of the source. It always decreases faster than  $(f/f_{\rm c})^{-1}$.  As the total GW energy must be finite, the  integral $\int_{0} ^\infty {\dot h}^2 (t) dt = \int _0 ^\infty f^2 {\tilde h}^2 (f) df$ yields an asymptotic bound of $g \propto (f/f_{\rm c})^{-\alpha_{\inf}}$ with $\alpha_{\inf} >3/2$. 

\begin{figure}
\centerline{\includegraphics[width=0.6\textwidth]{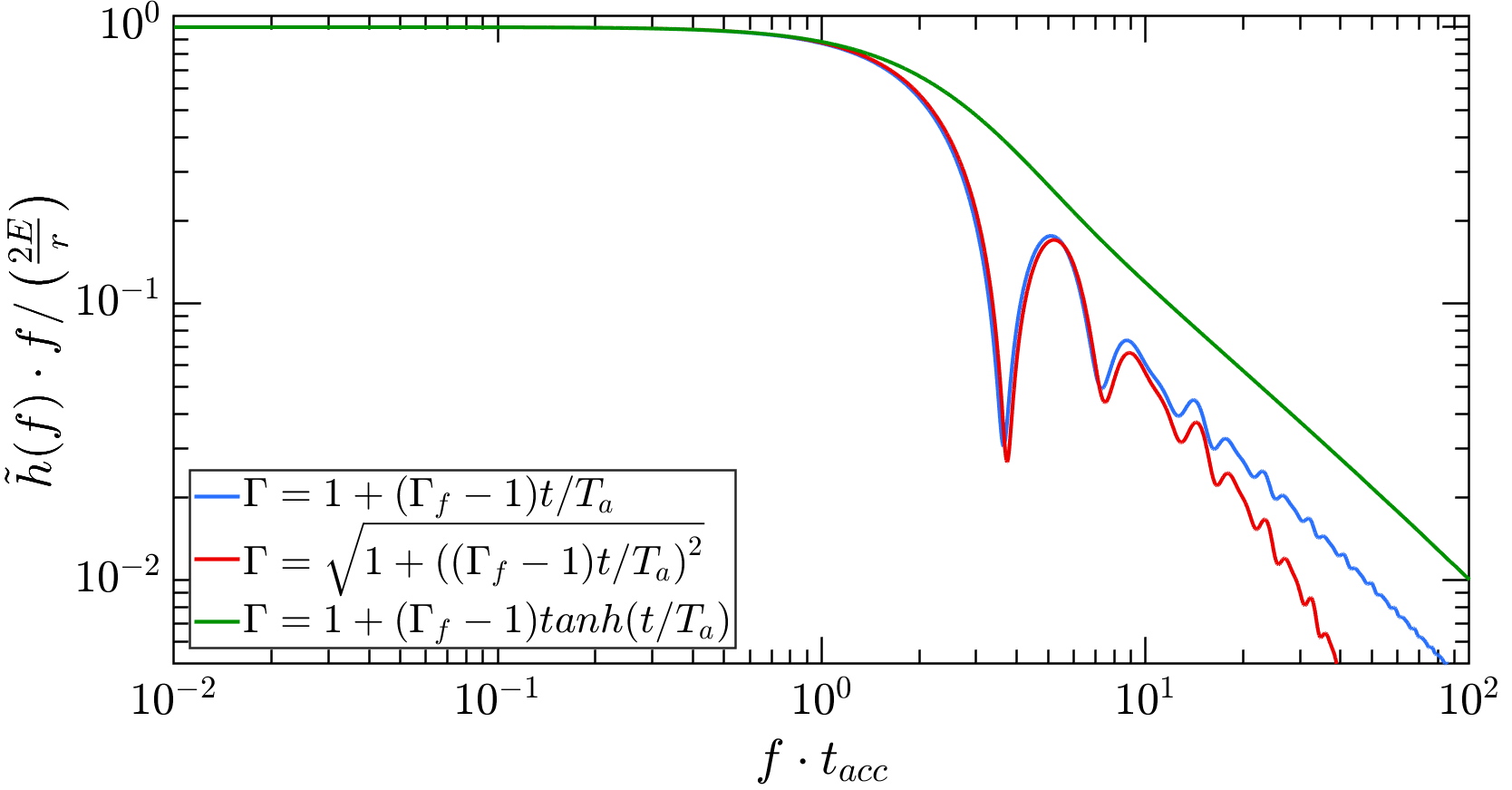}}
\centerline{\includegraphics[width=0.575\textwidth]{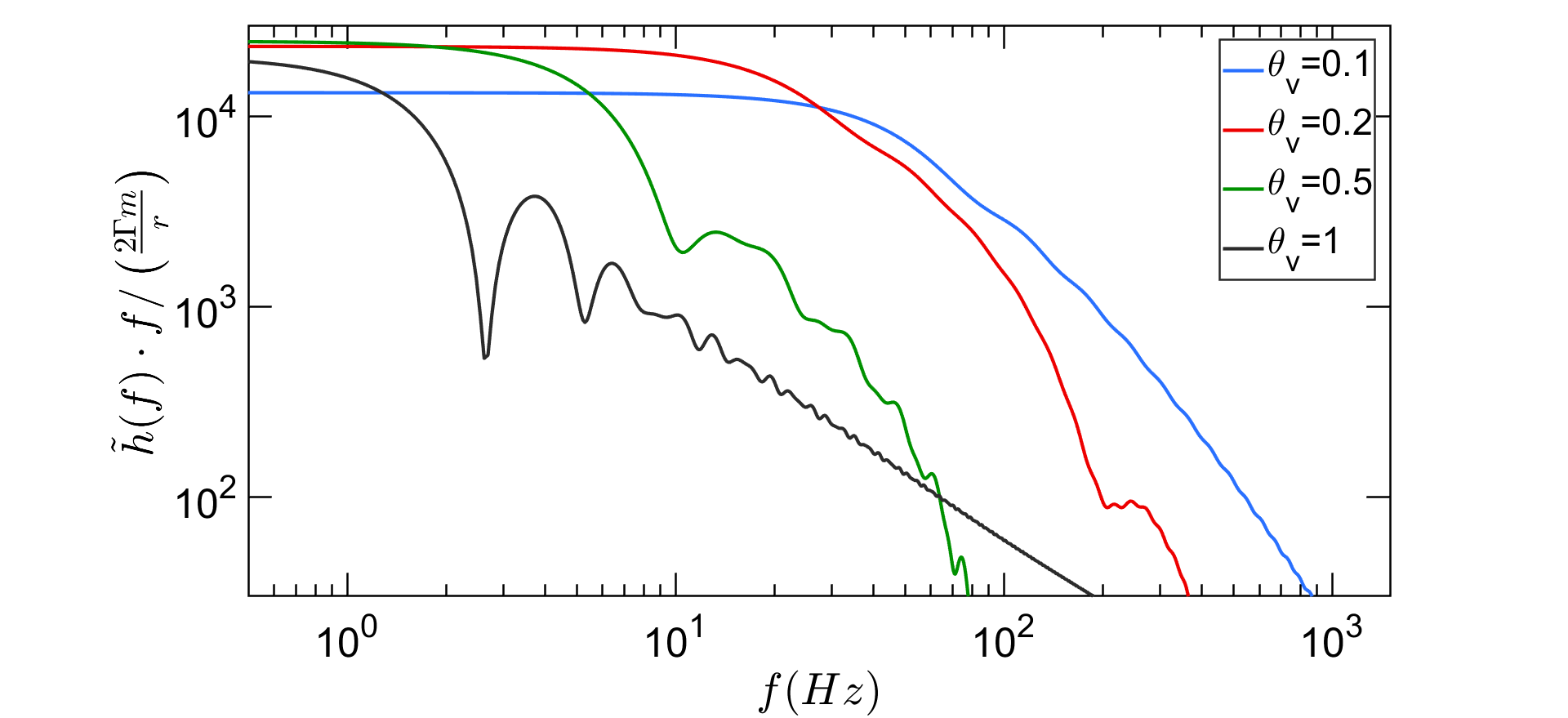}}
\caption{{\it top: }The Fourier transforms, $\tilde h$ of the GW signals from  jets with  three different acceleration models (with  $\Gamma_f = 100$, $\theta_{\rm j} = 0.1$, $\theta_{\rm v} = 0.9$). In all cases $t_{\rm inj}=0$. The time constant of the third model was chosen such that the high-frequency power laws would coincide. {\it Bottom:} The normalized Fourier waveform multiplied by the frequency for jets with the acceleration model $\Gamma \propto R$, based on the numerical code describe in \cite{ofek} for  
$\Gamma=100$ and $\theta_{\rm j}=0.1$. Below the crossover frequency,  $\tilde h(f) \cdot f$ is a constant. 
From ref.[7].}
\label{fig:h}
\end{figure} 

The {spectral density}, $S(f)\equiv \tilde h(f) \cdot \sqrt{f}$, is used to characterize the  sensitivity of GW  detectors. 
\begin{equation}
    \label{eq:crossover}
    S(f,\theta_{\rm v}) =  \begin{cases}  {h_0(\theta_{\rm v})}/{f^{1/2}} , & f \leq f_{\rm c}  
    \\ {h_0(\theta_{\rm v}) {g(f/f_{\rm c}) f^{1/2}} / f_{\rm c} } ,\ \ & f \geq f_{\rm c}  \ . \end{cases}
\end{equation}
The spectral density at the crossover frequency,  $S(f_{\rm c}) = {h_0(\theta_{\rm v})}/{f_{\rm c}^{1/2}} $,  is critical to determine the detectability of the signal. The condition
\begin{equation}
S_{det} (f )< S(f_{\rm c} ) (f_{\rm c} /f)^{1/2} = h_0(\theta_{\rm v})/{f^{1/2}} \ \ \ {\rm for ~ some~ f,}
\label{eq:Sfc}
\end{equation}  
is a necessary  condition for detection. 
For a low-frequency detector (whose typical frequency range is below $f_{\rm c} $) this condition is sufficient. However, it is not sufficient for a high frequency detector as $S(f)$ decreases faster than $f^{-1/2}$ above $f_{\rm c}$.

A more detailed discussion of the Fourier transform, $\tilde h(f,\theta_{\rm v})$ and the spectral density is given in ref. [7]. As an example Fig. \ref{fig:h}
depicts $\tilde h$ for different jets.

\section{The Angular Structure} 

\begin{figure}[b]
\centerline{\includegraphics[width=0.25\textwidth]{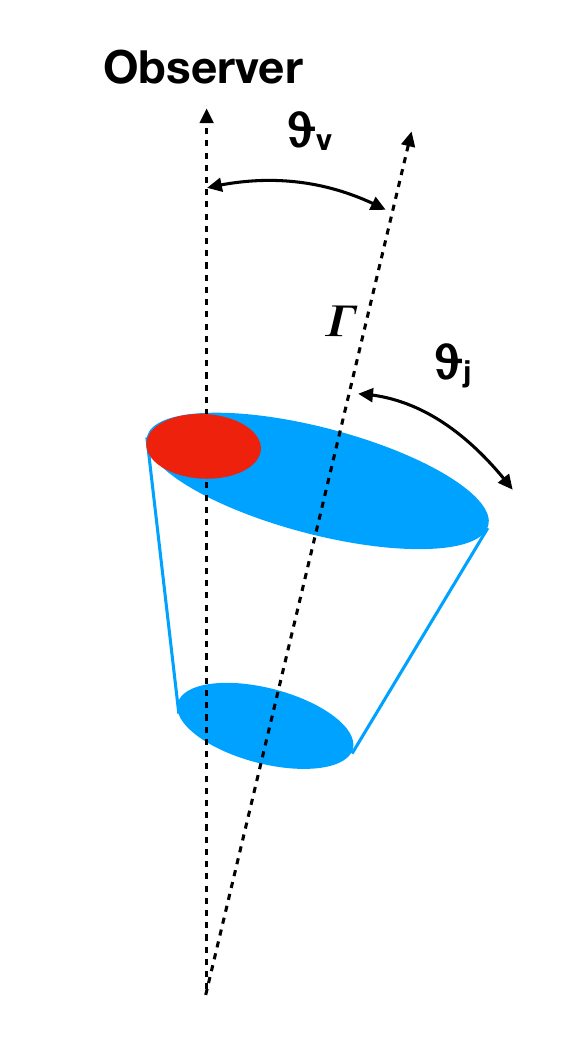}}
\caption{A schematic view  of the jet (blue) for $\theta_{\rm v} <\theta_{\rm j}$. Due to the symmetry, the contribution to the GW amplitude of the  part of the jet that is spherically symmetric around the observer
(shown in red ) vanishes. The amplitude from partial rings with $\theta>\theta_{\rm j}-\theta_{\rm v}$,   is reduced compared to the amplitude of a point-particle with the same energy and angle to the observer.
The jet is  symmetric under the transformation $\phi \rightarrow -\phi$: hence, the metric perturbation component $h_{\rm x}$ vanishes identically. From ref.[7].}
\label{fig:rings}
\end{figure}

To explore the effect of the angular width of the jet  we consider a thin spherical shell  ejected simultaneously and accelerated instantaneously.  
The shell has a half opening angle $\theta_{\rm j}$ and its center is moving at an angle $\theta_{\rm v}$ relative to the observer.  If  $\theta_{\rm j} \lesssim \Gamma^{-1}$ the signal converges to the point-particle limit. So in the rest we consider the case of  $\Gamma^{-1} \ll \theta_{\rm j}$. Otherwise 
cancellations between different parts of the shell become significant and the signal becomes weaker with a broader  anti-beaming region around the axis. 

We define the observer's line of sight to the emitting source as the $z$ axis of our coordinate system. The coordinates $\theta$ and $\phi$ are defined in the observer's coordinate system in the usual manner (see Fig. \ref{fig:schematic}). The direction of the center of the jet is $(\theta,\phi)=(\theta_{\rm v} , 0)$ in the observer's frame of reference.  
The axial symmetry implies a symmetry under the transformation $\phi \rightarrow -\phi$. Therefore, the metric perturbation $h_{\rm x}$ (which is now summed over the shell) vanishes identically (see Eq. \ref{eq:phase}, and Fig. \ref{fig:rings}). In the following, we simply denote $h=h_{\rm +}$, the only non-vanishing component of the metric perturbation tensor.

Integrating over the shell we find:
\begin{equation}
\label{eq:h_int}
    h_(\theta_{\rm v} , \theta_{\rm j}) =\frac{2\cE \beta^2}{r \Delta \Omega} \int_{|\theta_{\rm v}-\theta_{\rm j}|} ^{min(\theta_{\rm j}+\theta_{\rm v},\pi)}   \frac{\sin^3 \theta \cdot \sin 2\Delta \phi}{1-\beta \cos\theta}  d\theta \ , 
\end{equation}
where $\Delta \Omega \equiv 2 \pi (1-\cos\theta_{\rm j})$, the solid angle of the cap, and 
\begin{equation}
\label{eq:dphi}
    \Delta \phi \equiv  \cos^{-1} \left[ \frac{\cos \theta_{\rm j}-\cos \theta_{\rm v} \cos\theta}{\sin\theta_{\rm v} \sin\theta} \right]  \ .
\end{equation}

\begin{figure}
\centerline{\includegraphics[width=0.5\textwidth]{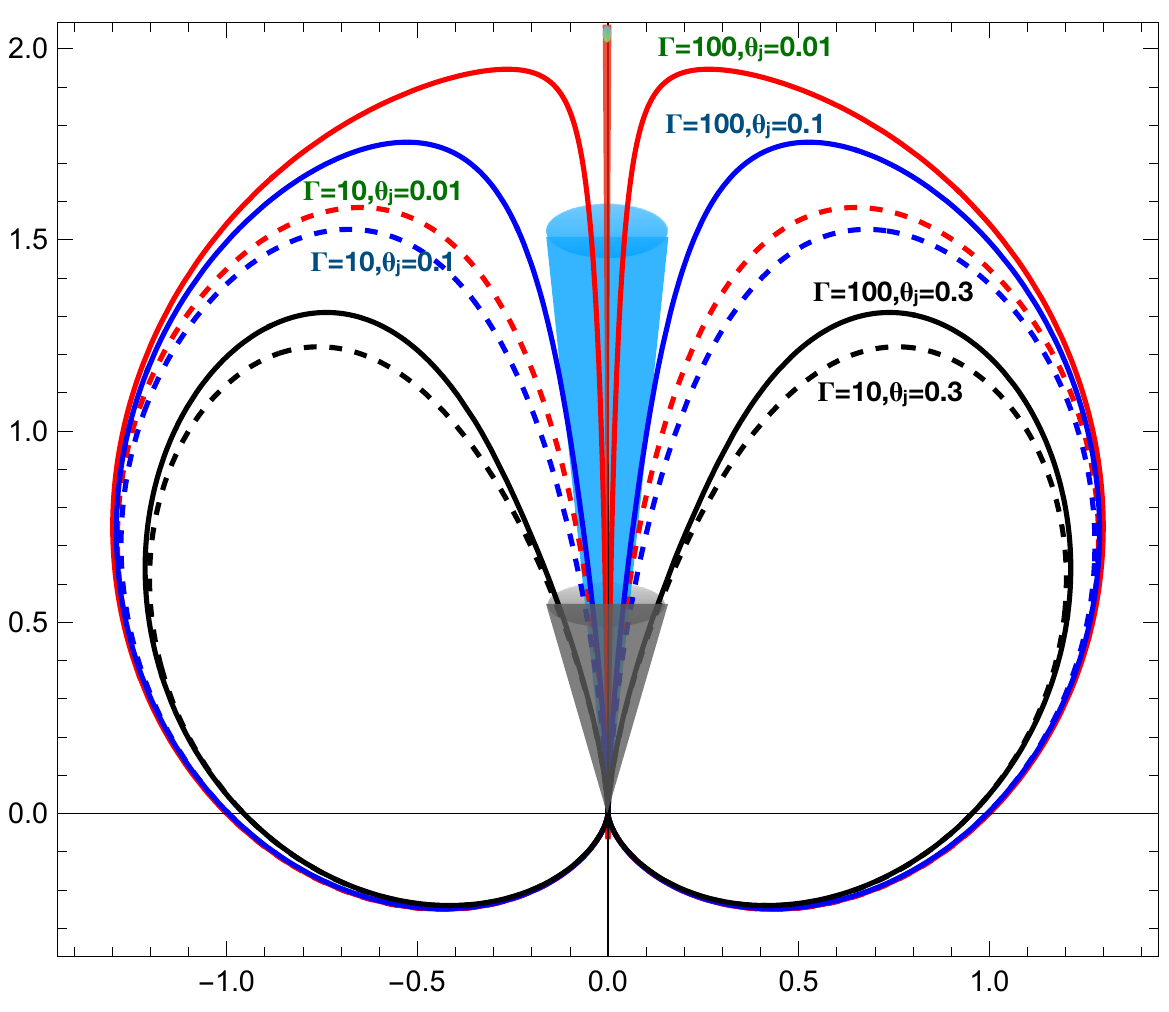}}
\caption{Antenna pattern for the GW amplitude, $h$, from an accelerating spherical shells with different opening angles $(\theta_{\rm j}=0.01,0.1,0.3$ - red, blue and black) and different Lorentz factors $(\Gamma=10,100$ - dashed and solid lines).  The anti-beaming region is  $ \approx 0.84 \ \theta_{\rm j}$. The narrow jet with $\theta_{\rm j}=0.01 \approx \Gamma^{-1}$ behaves like a point particle. For the larger opening angles ($\theta_{\rm j} >\Gamma^{-1}$) the pattern is almost independent of $\Gamma$. }
\label{theta_j_vs_pp}
\end{figure}

Figure \ref{theta_j_vs_pp} depicts the antenna pattern of $h(\theta_{\rm v},\theta_{\rm j})$ for different opening angles. This   angular behavior resembles the point-particle result, with a major difference: the anti-beaming region,
which was $ \Gamma^{-1}$ in the point-particle case, is now  $  \approx 0.84 \ \theta_{\rm j}$ 
and it is independent of $\Gamma$ (if $\Gamma^{-1}< \theta_{\rm j}$. If the jet is wide the signal is also weaker because of the cancellation of the signals from different parts of the shell. For $\theta_{\rm v} < \theta_{\rm j}$, 
only the outer region of the shell, with $\theta> \theta_{\rm j} - \theta_{\rm v}$,  
contributes. The effect is twofold: regions of the shell with $\theta<\theta_{\rm j}-\theta_{\rm v}$  
have a vanishing contribution to the amplitude, and even in the outer region, destructive interference between symmetric regions will reduce the GW amplitude. 
The maximal GW amplitude is now a function of $\theta_{\rm j}$ (compare with Eq. \ref{eq:hmax}). For small opening angles:
\begin{equation}
    h_{\rm max}(\theta_{\rm j}) \approx \frac{4 \cE}{r} (1-\frac{3}{4} \theta_{\rm j}) \ .
\end{equation}

Like in the case of a point source the GW signal of two sided jets with angular structure will be the sum of the two signals of the form given by Eq. \ref{eq:h_int}. The amplitude of $h$ will be almost isotropic but now with wider anti-beamed regions along the jets directions. Again the energy flux will be predominantly in the forward directions of the two jets.

\section{Detectability}
\label{sec:detectability}
 
To estimate detectability, we compare the expected spectral density, $S(f)$, to the detector's sensitivity curve that is characterized by   $S_{\rm det}(f)$.  $S(f)$ is always a decreasing function of the frequency.   At low frequencies, namely for $f< f_{\rm c}$,   $S(f) \propto f^{-1/2}$. It decreases faster for $f> f_{\rm c}$.  A typical  low-frequency detector will be most sensitive to a jet-GW  signal at the  lowest end of its frequency response.  Namely,  the lowest frequency below which $S_{\rm det}$ is steeper than $f^{-1/2}$.  At high-frequencies ($f>f_{\rm c}$)  $S \propto f^{-\alpha}$  with $\alpha > 1/2$.  A high-frequency detector is most sensitive for this signal at its lowest frequency below which $S_{\rm det}$ is steeper than $f^{-\alpha}$. 

Like almost any  GW source involving a relativistic motion, the maximal amplitude of the  jet GW is of order 
\begin{equation}
h \approx \frac{ 4 G \cE }{ c^4 r}\approx  10^{-24} \bigg(\frac{\cE}{10^{51} ~{\rm erg}}\bigg)~\bigg(\frac{100 {\rm Mpc}}{r} \bigg) \ . \label{eq:magnitude}
\end{equation}
$h= 10^{-24}$  is lower than the sensitivity of current GW detectors. Additionally, as we discuss later, these detectors don't operate  in the  deciHz frequency range that is relevant for most sources. However, we have to remember that  the sensitivity of current and planned detectors is much nearer to this signal  than the situation at the early 90ies when efforts to construct LIGO and Virgo begun. 

For a one-sided jet this estimate is valid for an observer that is at optimal angle, namely at $\theta_{\rm v} \approx \theta_{\rm j}$. For two-sided jets, this estimate is valid for most  observers apart from those along the jets ( $\theta_{\rm v} <  \theta_{\rm j}$).   Different observers may  observer different characteristic crossover frequencies (see Eq. \ref{eq.to}). In the following we will focus on observers that are in the forward direction (but not within the anti-beaming cones). 


\section{Potential Sources}
Detection horizons can be obtained by comparing the crossover frequency and amplitude of potential  sources with the characteristics frequencies and noise amplitudes of several detectors. Given the uncertainties in the signals and in properties of planned detectors this simple approach is  sufficient at this stage. 
Table 1.  summarizes the basic features of the different sources and  their detection horizon (assuming the high end of the energy range, a viewing angle near the jet and the most suitable frequency) with different GW detectors.  

\begin{table}[h]
\tbl{Horizon distances of several  current and proposed detectors for different sources.  Fig. \ref{fig:LIGO} describes the sensitivity curves of LVK, Einstern telscope, BBO and DECIGO.  For a moon type detector such as LWGA \cite{LWGA} or LSGA \cite{LSGA} I use sensitivity of $10^{-22} /\sqrt{Hz} $ at $10^{-3}Hz.$ The values are given using optimistic  values of  the energy (maximal) and the viewing angle (just outside the anti-beaming cone.}
{\begin{tabular}{@{}ccccccccc@{}} 
\toprule & Energy  & Frequency & Typical & LVK & Einstien & BBO  & DECIGO & Moon\cite{LWGA,LSGA} \\ 
Source &   ${\cal E}$ (erg) & $f_{\rm c}$ (Hz) &Distance  &   & Telescope & & & detectors \\
\colrule
LGRB & $10^{50-52}$ & 0.1-0.01 & 6 Gpc& -  & -  & Gpc  &  500 Mpc & 30 Mpc \\ 
sGRB & $10^{48-50}$ & 1-10  & 1 Gpc & -  & 5 Mpc  & 75 Mpc  &  50 Mpc & 3 Mpc\\
Sn Jet  & $10^{50-51}$ &  0.1-0.01  & 100 Mpc &-  & -  & 100 Mpc  &  50 Mpc & 3 Mpc \\
SGR & $10^{46-47}$  & 0.001  &10 kpc & 2.5 kpc  & 25 kpc  & 750 kpc  &  500 kpc & 10 kpc  \\
\botrule
\end{tabular}}\begin{tabnote}\end{tabnote}\label{ra_tbl1}
\end{table}

\begin{figure}
\centerline{\includegraphics[width=0.6\textwidth]{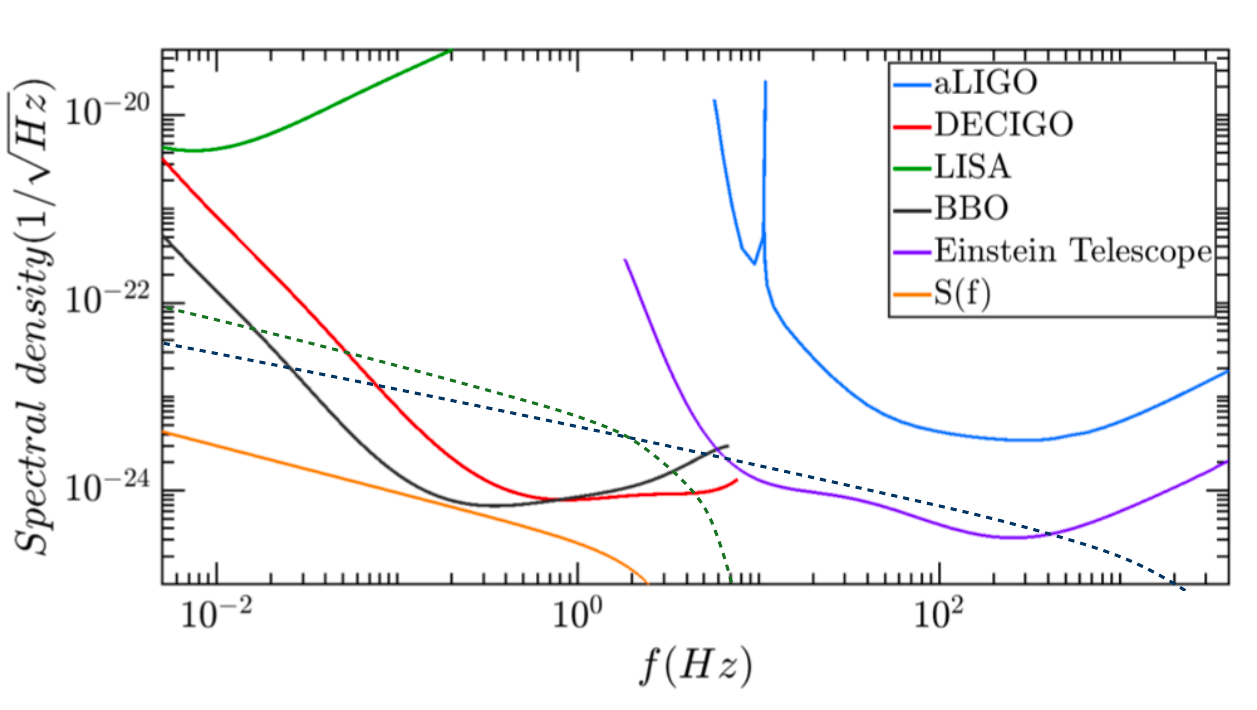}}
\caption{The estimated spectral density for a fiducial model\cite{Leiderschneider} of GW170817, $S(f)$, compared with the sensitivity thresholds of GW detectors taken from http://gwplotter.com.\cite{Moore_2014} The dotted red line shows the GW emission from a CCSN jet with $10^{51}$erg at 20 Mpc. The dotted black line corresponds to a Galactic SGR flare of $10^{47}$ erg at 10kpc.  Adapted with modifications from ref. [7]. }
\label{fig:LIGO}
\end{figure}

\subsection{GRBs}
GRB jets that have motivated these ideas\cite{segalis,piran2002} seems to be the most natural sources for these kind of GWs. These powerful ultra-relativistic jets carry up to $10^{52}$ erg (for long GRBs) and $10^{50}$ erg (for short ones). Given the numerous other differences between the two populations (long GRBs are typically detected from larger distances and they  have significantly larger  $t_{\rm inj}$) we consider each subgroup independently. The crossover frequency is dominated by $t_{\rm inj}$ for both long and short GRBs. Thus, $f_{\rm c}=f_{\it{l}}= 0.1-0.01$ Hz  for the long and $f_{\rm c}=f_{\it{s}}=1-10$ Hz for short GRBs.  This frequency range puts the events below the frequency limits of current LIGO-Virgo-Kagra,  but around the capability of the planned Big Bang Observer (BBO) \cite{BBO} and DECIGO \cite{Decigo}.

\subsubsection{Long GRBs}
Long GRBs (LGRBs) are associated with Collapsars - collapsing massive stars. Their typical isotropic equivalent energies range from $10^{51-54}$ ergs. When taking into account beaming, which is typically of order $\theta_{\rm j} \approx 0.1-0.2$ rad (but highly uncertain) this corresponds to ${\cal E} \approx  10^{50-52}$ erg. 
With typical distances of a few Gpc the GW amplitude from a regular LGRBs is quite low $\sim 10^{-26}$.
The duration of LGRBs ranges from 2 sec to a few hundred seconds, with typical duration of $\sim 20$ sec.  

\begin{figure}
\centerline{\includegraphics[width=0.6\textwidth]{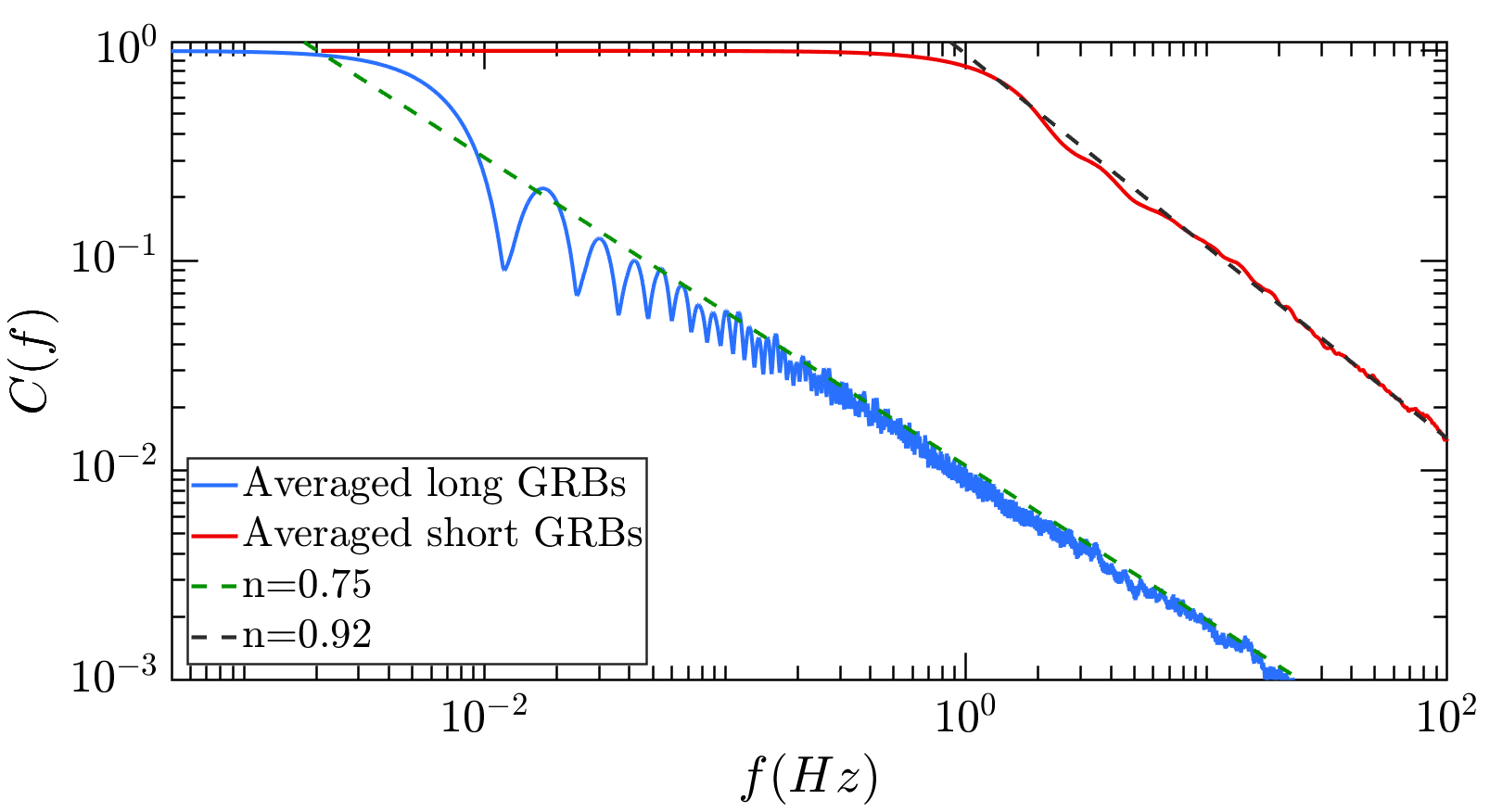}}
\caption{The averaged Fourier transform of BATSE's long GRBs \cite{Beloborodov}, vs. that of BATSE's TTE short GRB catalogue \cite{Leiderschneider}. Power law fits for are shown in dashed lines. From ref. [7]. }
\label{fig:tte}
\end{figure}

The average power spectrum of the bursts' $\gamma$-rays light curve shows a crossover frequency of $\approx 10^{-2} Hz$ \cite{Beloborodov}.  This crossover frequency reflects the overall duration of the bursts. If the duration of the $\gamma$-rays   reflects the activity of the central engine (as in the case of internal shocks \cite{kobayashi}) this implies that $t_{\rm inj}$  determines the dominant crossover frequency. This frequency is  below what can achieved by any detector built on Earth. It is at the upper end of  LISA's\footnote{\url{https://www.elisascience.org/}}  
bandwidth. However, the expected signal is about thousand times weaker than  LISA's  planned sensitivity (see Fig. \ref{fig:LIGO}). It falls nicely within the band of the proposed BBO  and DEFIGO. However, a regular LGRB at a few Gpc is too far to be detected with  the sensitivity reach of either  BBO  or DECIGO.  

\subsubsection{Short GRBs and GRB 170817A}
Short GRBs are associated with binary neutron star mergers \cite{eichler89}, as demonstrated recently with the observations of GRB 170817A and the associated GW signal GW 170817 \cite{Abbott2017}. With typical energies of $10^{48-50}$ erg, sGRBs are much weaker than LGRBs. Correspondingly they are observed from nearer distances, typically of order 1 Gpc, with the nearest ones at distances of a few hundred Mpc (170817A was much nearer than average). Typical sGRB duration is less than 2 sec. The expected crossover frequency (which is again, presumable, is dominated by $t_{\rm inj}$ and not by $t_{\rm acc}$) is of order 0.5-5 Hz. This is below current terrestrial detectors, it is too low even for  the planned frequency band of more advanced detectors like the Einstein telescope\footnote{\url{http://www.et-gw.eu/}}.  
Once more it falls within the planned frequency range  of BBO and DECIGO. Again a typical sGRB at a few hundred Mpc is too far to be detected. 

The exceptional sGRB,  GRB 170817A, was sufficiently nearby that it could have been marginally detected with these advanced detectors.  The properties of this well observed event are well known:  $\cE  \approx 10^{50} $erg,  $\theta_{\rm v}  \approx 20^o$, $\theta_{\rm j}\approx 5^o$ \cite{Nakar2020}. 
Other parameters, and in particular $t_{\rm acc} $ and $t_{\rm inj}$ that are most relevant for our analysis, are unknown. The injection duration $t_{\rm inj}$, is not necessarily related to the duration of the observed $\gamma$-rays in this event, as those arose from a cocoon shock breakout \cite{Kasliwal17,Gottlieb18}.  In the following we assume that $t_{\rm acc}< t_{\rm inj} \approx   1$ sec. 
$\Gamma$, is also unknown  but  it doesn't factor into the result as we expect $\Gamma^{-1} < \theta_{\rm j} ~ 0.1$ rad.  Given the viewing angle and the jet angle, GRB 170817A was also ideally positioned in terms of the strength of the GW signal from its jet. That is, we were not within the anti-beamed jet's cone but not too far from it either. 
The jet-GW of GRB 170717A signal was not  detected. This is not surprising.
Fig. \ref{fig:LIGO}, depicts the spectral density of GRB170817A compared with the sensitivity thresholds of GW detectors \cite{Moore_2014}. 
It is below the frequency band of both LIGO and Virgo and also well below their sensitivity. 
Jet-GW from  170817 would have been  detectable by BBO or DECIGO had such detector been operational at the time!

\subsection{Hidden jets in Core-Collapse Supernovae (CCSNe) and {\it low-luminosity} GRBs ({\it ll}GRBs)}

Shortly after the discovery of the first {\it ll}GRB 980415 (that was associated with SN98bw) it was suggested \cite{Kulkarni98,MacFadyen01,Tan01} 
that the emission arose from shock breakout following an energetic jet that was choked deep in the accompanying star. Later on it was realized  that, while the detection rate of {\it ll}GRBs is much lower than that of regular long ones, their actual rate is orders of magnitude larger \cite{Soderberg06,Bromberg11}. The detection rate is low because, given their low luminosity, they are detected only from relatively short distances. 
More recently,   it was shown\cite{Piran17} that a significant fraction of  CCSNe  that are not necessarily associated with GRBs contain  energetic ($\sim 10^{51}$ erg) choked relativistic jets. 
Such a jet deposits all its  energy into a cocoon. Upon breakout the cocoon material is observed as a  high velocity (0.1-0.2c) material that engulfs the supernova and can be observed within the first few days. Such signatures have been detected\cite{Mazzali00} as early as 1997  in SN 1997EF and in several other SNe since then. This suggestion was 
nicely confirmed with the exquisite observations of this high velocity material in  SN 2017iuk\cite{Izzo19,Nakar19}.  
It is supported by other evidence for jets in SNe \cite{totani2003,Mazzali2005,Maeda2008,Taubenberger2009,grichener2017}.

If such relativistic jets are associated with a significant fraction of CCSNe then, as the supernova rate is significantly larger than the GRB rate \cite{Piran_fireball}, we can expect  a significan number of relatively nearby jets that would be sources of jet-GWs. Comparing relativistic CCSNe Jets with GRB jets, we estimate a typical CCSN  $h$ to be a factor of 10  larger than the one from a typical long GRB.  
As we don't have a good clue on $t_{\rm inj}$ we assume that it is  of the same order as the one estimated in long GRB, namely  a few tens of seconds. Thus, the  crossover frequency would be around 0.01 Hz. 

If the event is accompanied by a {\it ll}GRB (arising from the shock breakout of the cocoon) then there will be almost simultaneous EM signal associated with the GW signal (typical delays could be of order of a dozen seconds). However, if the event is not associated with a {\it ll}GRB then the EM counterpart in the form of a supernova will emerge a few hours or even days after the GW signal. 

\subsection{Giant SGR Flares}
\label{sec:SGR}

Magnetars with magnetic fields larger than  $10^{15}$ Gauss are among the most surprising objects discovered \cite{Kouveliotou1999}. Magnetars eject giant flares while rearranging their whole magnetic field \cite{Eichler02}.  Such  flares have been observed  roughly once per ten years within the Milky Way.  The last one, from SGR1806-20 was observed in 2004\cite{Hurley2005,Mereghetti2005,Palmer2005,Terasawa2005}. It involved the  acceleration of  $\approx 10^{47}$ erg within a few milliseconds. The corresponding frequency is within the range of current GW detectors.  With a Galactic distance the signal might be detectable if the initial injection was beamed.  If  these flares involve a jet of such energy than they could be detected even with current GW detectors up to distances of a few kpc. Proposed more advanced detectors could observe them from the whole Galaxy or even from M31.

\subsection{Contribution to the GW background}
\label{sec:GWbackground}

The relativistic jets that arise from GRBs (both long and short) and hidden jets in SNe produce a  background of jet-GW waves at frequency range of $\sim  0.01-1  $Hz depending on the specific source.   Both long and short GRBs are rare and won't make a significant contribution to such a background. However,  CCSNe take place at a rate of about one per second in the observable Universe. If a significant fraction of SNe harbor energetic jets the time between two such cosmological events, a few seconds,  will be comparable to the characteristic time scale of the GW signals from these jets (assuming that the hidden jets in CCSNe are similar in nature to GRB jets).   Depending on the ratio of the time between events and the characteristic frequency of the jet-GW  signal we expect either a continuous background, as expected from the GW background from merging binary neutron stars,   or a pop-corn like signature, as expected for the GW background from merging binary black holes \cite{LIGOstochastic}.   With a typical cosmological distance  of a few Gpc the corresponding  amplitude of this jet-GW background is $h \approx 10^{-26} \cE/(10^{51} {\rm erg})$. With energy input of about $10^{41}$ erg in GW per SNe (see Eq. \ref{eq:egw}) and SN rate of about one per second in the Universe the resulting energy density of GW of this kind in the 0.1 Hz would be $\Omega_{\rm jet-GW}\approx 10^{-9}$.  Note, however, that there is an uncertainty of at least an order of magnitude due to the unclear $t_{\rm c}$ of this signal.

\section{Discussion}
\label{sec:discussion}
The acceleration of a jet to a relativistic velocity generates a unique memory type GW signal.  The signal is anti-beamed away from the direction of the jet with an anti-beaming angle of $\max(\Gamma^{-1},\theta_{\rm j}$).
Like typical relativistic GW sources, the amplitude is of order $G\cE/c^4 r $, corresponding to an amplitude of $h \approx 10^{-24}$ for a jet of $10^{51}$ erg at $100$ Mpc. 
At low frequencies the signal can be approximated as a step function.  This last feature is of course problematic, as it might be difficult to distinguish it from step function type noise sources. At higher frequencies the details of the signal lightcurve could reveal information about the acceleration and collimation  of the jet. 

Prospects for detection of these jet-GWs depend on construction of appropriate future detectors. Jets from different kinds of GRBs as well as hidden jets in CCSNe are expecte in the deciHz range. Hence most important are detectorslike the proposed BBO, DECIGO and several  moon detectors (e.g. LWGA, LSGA). Among those sources hidden jets in CCSNe are  the best candidates as they are most frequent and as their expected energy is comparable to those of LGRBs. 

As jet-GWs are anti-beamed we will most likely miss the $\gamma$-rays associated with GRBs from which we will observe these GWs. However, we expect to observe other associated wider signatures such as the afterglow or an accompanying shock breakout signal from the stellar envelope (in case of a LGRB) or from the merger's ejecta (in a sGRB). 
This  shock breakout signal is also be the most likely EM signal associated temporarily with a jet-GW arising from hidden jet in CCSN. In this case the actual SN signal may appear hours later. 

Table 1 summarizes (under somewhat optimistic assumptions) the detection horizon of different events. Given the large uncertainties I didn't attempt to incorporate possible estimates of detection rates as these   will be highly uncertain. Detection prospects with current detectors, LIGO, Virgo and Kagra that are operating at frequencies above 10 Hz are slim, unless giant SGR flare involve jets. In that case we could detect a Galactic event even with LVK. Otherwise we will have to resort of future planned detectors. What is most important is to remember that these signals are memory type. Usually we are not looking for such astrophysical signals. We should make sure to expand GW searches to include such signatures in all future runs.  
A detection of Jet-GWs would open  a new window on jet engines and would reveal features of jet acceleration in the vicinity of black holes that are impossible to explore in any other way. 

\begin{acknowledgments}
The research was supported by an advanced ERC grant TReX. 
\end{acknowledgments}

\bibliographystyle{ws-rv-van}
%

\end{document}